\documentclass[aps,prd,reprint,showpacs,amsmath,amssymb,superscriptaddress]{revtex4-1}

\usepackage{graphicx}

\newcommand{\bA}{\boldsymbol{A}}

\newcommand{\br}{\boldsymbol{r}}
\newcommand{\bx}{\boldsymbol{x}}

\newcommand{\by}{\boldsymbol{y}}

\newcommand{\uA}{\textup{A}}

\newcommand{\uF}{\textup{F}}
\newcommand{\uH}{\textup{CS}} 
\newcommand{\uS}{\textup{LL}}

\newcommand{\Sc}{\mathcal{S}}
\newcommand{\R}{\mathbb{R}}
\newcommand{\Z}{\mathbb{Z}}

\newtheorem{thm}{Theorem}
\newtheorem{lem}[thm]{Lemma}

\begin{document}

\title{Local exclusion principle for identical particles\\obeying intermediate and fractional statistics}

\author{Douglas Lundholm}
\altaffiliation{Work partly done while visiting FIM, ETH Z\"urich (D.L.), and Institutes Mittag-Leffler and Henri Poincar\'e (both authors).}
\affiliation{Department of Mathematics, KTH Royal Institute of Technology, SE-10044 Stockholm, Sweden}
\author{Jan Philip Solovej}
\altaffiliation{Work partly done while visiting FIM, ETH Z\"urich (D.L.), and Institutes Mittag-Leffler and Henri Poincar\'e (both authors).}
\affiliation{Department of Mathematical Sciences, University of Copenhagen, Universitetsparken 5, DK-2100 Copenhagen \O, Denmark}

\begin{abstract}
	A local exclusion principle is observed for 
	identical particles obeying intermediate/fractional exchange
	statistics in one and two dimensions, 
	leading to bounds for the kinetic energy in terms of the density.
	This has implications for models of 
	Lieb-Liniger and Calogero-Sutherland type,
	and implies a non-trivial lower bound for the energy of the anyon gas 
	whenever the statistics parameter is an odd numerator fraction.
	We discuss whether this is actually a necessary requirement.
\end{abstract}


\maketitle

\section{Introduction}

	The majority of interesting phenomena 
	in many-body quantum mechanics are in some way
	associated to the fundamental concept of identical particles and statistics.
	Elementary identical particles  
	in three spatial dimensions are either 
	bosons, obeying Bose-Einstein statistics,
	or fermions, obeying Fermi-Dirac statistics.
	The former are usually represented using wave functions 
	which are symmetric under
	particle permutations, while the latter 
	implement Pauli's exclusion principle by exhibiting 
	total anti-symmetry under particle interchange.
	On the other hand, for point particles living
	in one and two dimensions there 
	are logical possibilities
	different from bosons and fermions,
	so-called intermediate or fractional statistics 
	\cite{Streater-Wilde:70,Leinaas-Myrheim:77,Goldin-Menikoff-Sharp:81 & Wilczek:82}.
	Although first regarded as 
	of purely academic interest
	--- filling the loopholes in the arguments leading to the two 
	standard permutation symmetries ---
	these have recently become 
	a reality in the laboratory, 
	with the advent of trapped bosonic gases \cite{ref:trapped-bosons}
	and quantum Hall physics \cite{ref:QHE},
	and thus the discoveries of effective models of particles
	(or quasi-particles) that seem to obey these generalized rules
	for identical particles and statistics.
	We refer to \cite{Myrheim:99,Khare:05,ref:reviews} 
	for extensive reviews on these topics.
	
	Although
	non-interacting bosons and fermions are well understood
	in terms of single-particle Hilbert spaces and operators,
	the same cannot be said about particles obeying these
	generalized interchange statistics.
	Namely, despite some effort in this direction 
	\cite{Haldane:91,Isakov:94},
	many-particle quantum states for 
	intermediate and fractional 
	exchange statistics 
	have in general not admitted a 
	simple description
	in terms of single-particle states 
	restricted by 
	some exclusion principle. 
	The reason for this difficulty is that the general symmetry 
	of the wave function
	under particle interchange is naturally modeled using
	pairwise or many-body interactions,
	hence leaving the much simpler realm of single-particle Hamiltonians
	(and also introducing other mathematical difficulties as well,
	already at the formulation of these models).
	
	As a different approach,
	we would like to stress in the following that 
	the effects of exclusion are also encoded in
	inequalities for many-particle energy forms, such as
	the Lieb-Thirring inequality \cite{Lieb-Thirring:75}.
	For the case of identical spinless fermions 
	in an external potential $V$
	in $d$-dimensional space,
	it states that there is a \emph{uniform} bound
	for the energy of a normalized $N$-particle state $\psi$:
	\begin{equation} \label{LT}
		\langle \psi, \hat{H}\psi \rangle
		\ge \ -\sum_{k=0}^{N-1} |\lambda_k|
		\ \ge \ -C_d \int |V_-(\bx)|^{1+d/2} \,d^d\bx,
	\end{equation}
	with the $N$-particle Hamiltonian operator
	$$
		\hat{H} = \hat{T}_0 + \hat{V} = \sum_{j=1}^N \left( -\frac{1}{2}\nabla_j^2 + V(\bx_j) \right),
	$$
	the conventions $\hbar = m = 1$, $V_\pm:=(V \pm |V|)/2$,  
	and a positive constant $C_d$.
	The inequality \eqref{LT}
	incorporates Pauli's exclusion principle via 
	the intermediate
	sum over the 
	negative energy levels $\lambda_k$ 
	of the one-particle Hamiltonian
	$\hat{h} = -\frac{1}{2}\nabla^2 + V(\bx)$. 
	It furthermore 
	incorporates the uncertainty principle,
	and is in fact equivalent to the kinetic energy inequality
	\begin{equation} \label{kinetic-LT}
		\langle \psi, \hat{T}_0 \psi \rangle
		\ \ge \ \frac{d \,(2/C_d)^{2/d}}{(d+2)^{1+2/d}} \int \rho(\bx)^{1+2/d} \,d^d\bx,
	\end{equation}
	involving the one-particle density $\rho$ of $\psi$; 
	normalized
	$\int \rho(\bx) \,d^d\bx = N$.
	In dimension $d=3$,
	the expression on the r.h.s. of \eqref{kinetic-LT} may be recognized
	as the kinetic energy approximation from Thomas-Fermi theory.
	It is in this case conjectured \cite{Lieb-Thirring:75} 
	that \eqref{kinetic-LT} holds
	with exactly the Thomas-Fermi expression on the right.
	The best known result is, however, smaller by a factor $(3/\pi^2)^{1/3}$
	\cite{Dolbeault-Laptev-Loss:08}.
	
	The bounds \eqref{LT} and \eqref{kinetic-LT}
	need to be weakened in the case of weaker exclusion.
	In the case that each single-particle state can be filled $q$ times
	(e.g. in models with $q$ spin states, 
	or cp. Gentile intermediate statistics \cite{Gentile:40-42})
	the r.h.s. of the inequalitites 
	\eqref{LT} resp. \eqref{kinetic-LT} are to be multi\-plied
	by $q$ resp. $q^{-2/d}$.
	Bosons can then be accommodated by $q=N$,
	yielding trivial bounds as $N \to \infty$.
	
	In this work we wish to report on 
	a new set of Lieb-Thirring-type
	inequalities for intermediate and fractional statistics,
	which follow from a corresponding \emph{local} version 
	of the exclusion principle,
	applicable to such interacting systems.
	Our approach is 
	very much inspired by the work \cite{Dyson-Lenard:67}
	of Dyson and Lenard (see also \cite{Dyson-Lenard:reviews}),
	who used only such a local form of the 
	Pauli principle to rigorously prove 
	the stability of ordinary fermionic matter in the bulk
	(the inequalities \eqref{LT} and \eqref{kinetic-LT}
	were subsequently invented by Lieb and Thirring
	to simplify their proof).
	Although the numerical constants resulting from our method are
	comparatively weak, we believe the forms of our bounds to be conceptually
	very useful, and as a result we also learn something
	non-trivial about the elusive anyon gas.
	
	Starting by recalling the models for intermediate and fractional
	statistics which we shall be concerned with here, we proceed by showing
	how a local form of the exclusion principle can be established for
	such statistics, leading to bounds for the kinetic energy in terms
	of the one-particle density $\rho(\bx)$.
	For clarity, we leave out some of the technical details,
	referring to the mathematical papers
	\cite{Lundholm-Solovej:anyon,Lundholm-Solovej:extended}, 
	and instead focus on general aspects of the procedure.
	With these preparations, we consider 
	the problem of
	determining the ground state energy for a large number $N$ of anyons
	in a harmonic oscillator potential, 
	and can conclude that the energy grows like $N^{3/2}$
	under the assumption that 
	the anyonic statistics phase is an odd numerator rational 
	multiple of $\pi$.
	In the final section we discuss a 
	structural difference between such odd and even 
	numerator fractions 
	using a class of 
	trial states which are related to the Read-Rezayi states for the 
	fractional quantum Hall effect.

\section{Identical particles in one and two dimensions}

	We recall three well-established models 
	for intermediate and fractional exchange statistics
	for scalar non-relativistic quantum mechanical particles
	in one or two spatial dimensions.
	As mentioned in the introduction, there are by now
	a number of standard references for their background and derivations,
	which we will accordingly skip here.
	We will mainly follow the notation in \cite{Myrheim:99}, with 
	technical details addressed in \cite{Lundholm-Solovej:extended}.
	
	Identical particles in 2D, anyons, have the possibility to
	pick up an arbitrary but fixed phase $e^{i\alpha\pi}$
	upon continuous simple interchange of two particles
	\cite{Leinaas-Myrheim:77,Goldin-Menikoff-Sharp:81 & Wilczek:82}.
	A standard way to model such (abelian) anyons,
	in the so-called magnetic gauge, 
	is by means of bosons in $\R^2$ together with a statistical 
	magnetic interaction given by the vector potential
	$$
		\bA_j = \alpha \sum_{k \neq j} \frac{(\bx_j - \bx_k)I}{|\bx_j - \bx_k|^2},
		\qquad \alpha \in \R \pmod 2,
	$$
	where $\bx I$ denotes a $90^\circ$ counter-clockwise rotation 
	of the vector $\bx$.
	This attaches to every particle 
	an Aharonov-Bohm point flux
	of strength $2\pi\alpha$, 
	felt by all the other particles.
	The kinetic energy for $N$ such particles is thus given by
	$T_{\uA} := \langle\psi, \hat{T}_{\uA} \psi\rangle$, 
	\begin{equation} \label{kinetic-energy-2D}
		\hat{T}_{\uA} := \frac{1}{2} \sum_{j=1}^N D_j^2,
	\end{equation}
	where $D_j = -i\nabla_j + \bA_j$,
	and the wave function
	$\psi$ is represented as a completely symmetric square-integrable
	function on $(\R^2)^N$.
	The case $\alpha=0$ then corresponds to bosons, 
	and $\alpha=1$ to fermions.

	The case of 
	identical particles confined to move in 
	only one spatial dimension
	is special and in some sense degenerate, 
	since particles cannot be interchanged continuously without colliding.
	In quantum mechanics this necessitates some choice of 
	boundary conditions for the wave function at the collision points.
	It turns out that,
	depending on which approach one takes to quantization 
	\cite{Leinaas-Myrheim:77,1D-refs,Myrheim:99},
	identical particles in 1D can again be modeled as bosons,
	i.e. wave functions symmetric under the flip 
	$r \mapsto -r$ of any two relative particle coordinates
	$r := x_j - x_k$,
	together with a local interaction potential, singular at $r=0$
	and either of the form $\delta(r)$ or $1/r^2$.
	We write 
	\begin{equation} \label{statistics-potentials}
		V_{\uS}(r) := 2\eta \delta(r), \qquad 
		V_{\uH}(r) := \frac{\alpha(\alpha-1)}{r^2}, 
	\end{equation}
	with statistics parameters $\eta,\alpha \in \R$,
	for the respective models 
	resulting from a 
	Schr\"odinger- resp. Heisenberg-type approach to quantization.
	These statistics potentials
	correspond to the choices of boundary conditions
	for the wave function $\psi$ at the boundary $r=0$ 
	of the configuration space
	$$
		\frac{\partial \psi}{\partial r} = \eta \psi, \ \ \text{at $r=0^+$},
		\quad \text{resp.} \quad 
		\psi(r) \sim r^\alpha, \ \ \text{as $r \to 0^+$}.
	$$
	Here $\eta=0$ resp. $\alpha=0$ represent bosons
	(Neumann b.c.) while $\eta=+\infty$ resp. $\alpha=1$
	represent fermions (Dirichlet or analytically vanishing b.c.;
	see \cite{Lundholm-Solovej:extended}). 
	Suggested by such pairwise boundary conditions, 
	one may define \cite{Note:defs-SH} 
	the total kinetic energy for a normalized
	completely symmetric wave function $\psi$
	describing $N$ identical particles on the full real line $\R$ to be
	$T_{\uS/\uH} := \langle\psi, \hat{T}_{\uS/\uH} \psi\rangle$
	where 
	\begin{equation} \label{kinetic-energy-1D}
		\hat{T}_{\uS/\uH} := -\frac{1}{2} \sum_{j=1}^N \frac{\partial^2}{\partial x_j^2} 
			+ \sum_{1\le j<k \le N} V_{\uS/\uH}(x_j - x_k).
	\end{equation}
	In other words, the $\uS$ 
	case in our notation is nothing but the 
	Lieb-Liniger model for one-dimensional bosons
	with pairwise Dirac delta interactions \cite{Lieb-Liniger:63},
	while the $\uH$ 
	case corresponds to the homogeneous part
	of the Calogero-Sutherland model with inverse-square interactions
	\cite{Calogero:69 & Sutherland:71}.
	It is well-known that 
	these models can describe a continuous interpolation between the 
	properties of bosons and fermions
	for certain ranges of the statistics parameters.
	For the following results 
	we will restrict to 
	$\eta \ge 0$ (Lieb-Liniger 
	type intermediate statistics) 
	and $\alpha \ge 1$ (Calogero-Sutherland 
	type `superfermions')
	for which the statistics potentials 
	\eqref{statistics-potentials} are nonnegative, i.e. repulsive.

	In all of the above cases, the one-particle density $\rho(\bx)$
	is defined s.t. the expected number of particles on a local
	region $Q$ of space (typically a $d$-dimensional
	cube in the following) equals
	$$
		\int_Q \rho(\bx) \,d\bx = \sum_{j=1}^N \int_{\R^{dN}} |\psi|^2 \,\chi_Q(\bx_j) \,dx,
	$$
	where $\chi_Q \equiv 1$ on $Q$ and $\chi_Q \equiv 0$ 
	on the complement $Q^c$.
	In particular, $\int_{\R^d} \rho = N$.
	Similarly,
	it is useful to be able to speak about the expected kinetic
	energy of a wave function on a local region. 
	For fermions or bosons we naturally define this quantity to be
	$$
		T^Q_0 := \frac{1}{2} \sum_{j=1}^N \int_{\R^{dN}} |\nabla_j \psi|^2 \,\chi_Q(\bx_j) \,dx.
	$$
	Analogously for anyons,
	$$
		T^Q_{\uA} := \frac{1}{2} \sum_{j=1}^N \int_{\R^{2N}} 
		|D_j \psi|^2 \,\chi_Q(\bx_j) \,dx,
	$$
	and for 1D intermediate statistics, 
	$T^Q_{\uS/\uH} := $
	$$
		\frac{1}{2} \sum_{j=1}^N \int_{\R^N} \!\!
		\left( |\partial_j \psi|^2 + \sum_{k \neq j} 
		V_{\uS/\uH}(x_j - x_k) |\psi|^2 \right) \!\!
		\chi_Q(x_j) \,dx.
	$$
	Note that if the full space $\R^d$ 
	has been partitioned into a family of non-overlapping regions
	$\{Q_k\}$ then the total kinetic energy decomposes as
	$T_{0/\uS/\uH/\uA} = \sum_k T_{0/\uS/\uH/\uA}^{Q_k}$.
	We will furthermore write 
	$T_{\uF} = T_0$ to denote the free kinetic energy
	for the particular case of fermions in $\R^3$,
	i.e. totally antisymmetric $\psi$.

\section{Local exclusion}

	The starting point for our energy bounds
	will be the following local consequence of the
	Pauli exclusion principle for fermions, which was used by
	Dyson and Lenard in their proof of stability of matter \cite{Dyson-Lenard:67}:
	Let $\psi$ be a wave function of $n$ spinless fermions in $\R^3$,
	i.e. anti-symmetric w.r.t. every pair of particle indices, 
	and let $Q$ be a cube of side length $l$.
	Then, 
	for the contribution to the free kinetic energy
	with all particles in $Q$,
	\begin{equation} \label{local_exclusion_F}
		\frac{1}{2} \int_{Q^n} \sum_{j=1}^n |\nabla_j \psi|^2 \,dx
		\ \ge \ (n-1) \frac{\xi_{\uF}^2}{l^2} \int_{Q^n} |\psi|^2 \,dx,
	\end{equation}
	where $\xi_{\uF} = \pi/\sqrt{2}$.
	In other words, due to the Pauli principle,
	the energy is nonzero for $n \ge 2$
	and grows at least linearly with $n$
	(indeed linearity proves to be sufficient; 
	cp. also \cite{Dyson-Lenard:reviews}).
	In \cite{Dyson-Lenard:67}, $Q$ was replaced by a ball of radius $l$
	and $\sqrt{2}\xi_{\uF}$ by the smallest positive root of the equation
	$(d^2/d\xi^2)(\sin \xi/\xi) = 0$.
	The inequality \eqref{local_exclusion_F} follows by expanding $\psi$ in 
	the eigenfunctions of the Neumann Laplacian on $Q$, 
	or by the pairwise method 
	below at the cost of a slightly
	weaker constant $\xi_{\uF}$.

\begin{figure}[t]
		\centering
		\includegraphics[scale=0.35]{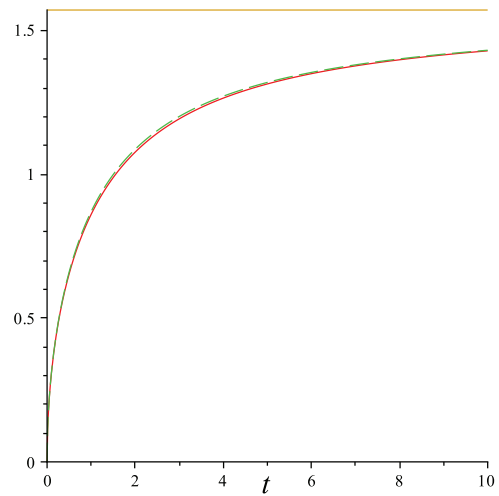}
		\caption{Plot of $\xi_{\uS}(t)$ (red solid) and 
		$\arctan \sqrt{t + 4t^2/\pi^2}$ (green dashed) 
		as a function of $t \ge 0$.}
		\label{fig:xiS}
	\end{figure}

\begin{figure}[t]
		\centering
		\includegraphics[scale=0.34]{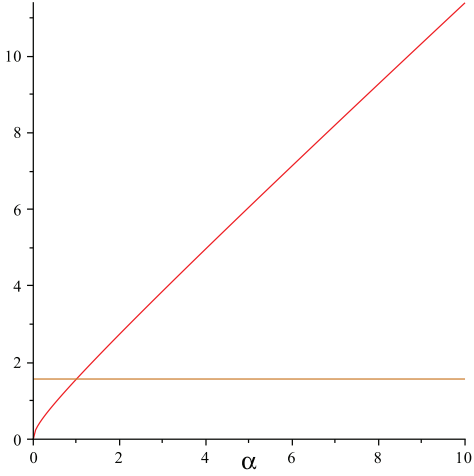}
		\caption{Plot of $\xi_{\uH}(\alpha)$ 
		as a function of $\alpha \ge 0$.}
		\label{fig:xiH}
	\end{figure}

	Now, for the 1D case we introduce 
	$\xi_{\uS}(\eta l)$ resp. $\xi_{\uH}(\alpha)$ 
	to be the smallest positive solutions of 
	$\xi \tan \xi = \eta l$, resp. 
	$(d/d\xi)(\xi^{1/2} J(\xi))=0$,
	where $J$ is the Bessel function of order $\alpha-1/2$.
	These $\xi_{\uS/\uH}$ arise as quantization conditions for the 
	wave function upon considering the Neumann problems 
	\begin{equation} \label{Neumann_problems}
		(-\partial_r^2 + V_{\uS/\uH}(r))\psi = \lambda \psi, \quad
		\partial_r \psi|_{r = \pm l} = 0
	\end{equation}
	in the pairwise relative coordinate $r$ on an interval $[-l,l]$, 
	yielding a lowest bound for the energy $\lambda = \xi_{\uS/\uH}^2/l^2$.
	A good numerical approximation to $\xi_{\uS}$ is given by
	$\xi_{\uS}(t) \approx \arctan \sqrt{t+4t^2/\pi^2}$
	for all $t \ge 0$
	(see Fig. \ref{fig:xiS}), while
	we have $\xi_{\uH}(1) = \pi/2$ and, asymptotically,
	$\xi_{\uH}(\alpha) \sim \alpha$ as $\alpha \to \infty$
	(see Fig. \ref{fig:xiH}).
	
	In the case of anyons we define the expression
	\begin{equation} \label{anyon_constant}
		\xi_{\uA}(\alpha,n) 
		:= \min_{p \in \{0,1,\ldots,n-2\}} \min_{q \in \Z} 
			|(2p+1)\alpha - 2q|,
	\end{equation}
	which
	measures the fractionality
	of the parameter $\alpha$ and arises 
	in a bound for a local pairwise 
	magnetic operator, which is the 2D analog to \eqref{Neumann_problems} 
	(and defined on an annulus instead of an interval \cite{Lundholm-Solovej:anyon}).
	The absolute quantity which is being minimized in
	\eqref{anyon_constant} can be understood as 
	the magnetic gauge phase $(2p+1)\alpha\pi$ 
	arising from a pairwise interchange
	of two anyons --- with
	the odd integer $2p+1$ depending on the number $p$ of other
	anyons that can appear inside such a two-anyon interchange loop
	and the additional $+1$ stemming from the statistics flux 
	of the interchanging pair itself. 
	This is taken
	modulo the pairwise orbital angular momentum of the wave function which 
	is an even integer $-2q$
	due to the underlying bosonic symmetry. 
	Note that for bosons $\xi_{\uA}(\alpha=0,n) \equiv 0$
	while for fermions we have $\xi_{\uA}(\alpha=1,n) \equiv 1$
	for all $n$.
	
	We call the following observation a \emph{local exclusion principle} 
	for generalized exchange statistics since it implies that the local kinetic energy 
	is nonzero whenever we have more than one particle, 
	and hence that the particles cannot occupy the same single-particle state 
	(which on a local region would be the zero-energy ground state).
	
	\begin{lem}[Local exclusion principle]
		Given any finite interval $Q \subset \R$ of length $|Q|$,
		we have for $\eta \ge 0$ 
		\begin{equation} \label{local_exclusion_1S}
			\int_{Q^n} \bar{\psi} \,\hat{T}_{\uS} \psi \,dx
			\ \ge \ (n-1) \frac{\xi_{\uS}(\eta|Q|)^2}{|Q|^2} 
				\int_{Q^n} |\psi|^2 \,dx,
		\end{equation}
		and for $\alpha \ge 1$
		\begin{equation} \label{local_exclusion_1H}
			\int_{Q^n} \bar{\psi} \,\hat{T}_{\uH} \psi \,dx
			\ \ge \ (n-1) \frac{\xi_{\uH}(\alpha)^2}{|Q|^2} 
				\int_{Q^n} |\psi|^2 \,dx,
		\end{equation}
		while for a square $Q \subset \R^2$ with area $|Q|$
		and any $\alpha \in \R$ 
		\begin{equation} \label{local_exclusion_anyon}
			\frac{1}{2} \int_{Q^n} \sum_{j=1}^n |D_j \psi|^2 \,dx
			\ \ge \ (n-1) \frac{c\,\xi_{\uA}(\alpha,n)^2}{|Q|} 
				\int_{Q^n} |\psi|^2 \,dx,
		\end{equation}
		with $c = 0.056$.
		It then follows for the expected kinetic energy on 
		a $d$-dimensional cube $Q$ with volume $|Q|$ that 
		\begin{equation} \label{local_exclusion_density_all}
			T^Q_{\uS/\uH/\uA/\uF} 
			\ \ge \ \frac{\xi_{\uS/\uH/\uA/\uF}^2}{|Q|^{2/d}} 
				\left( \int_Q \rho(\bx) \,d\bx \ - 1 \right)_+,
		\end{equation}
		where $\xi_{\uS/\uH/\uA/\uF}$ here stands for 
		$\xi_{\uS}(\eta|Q|)$,
		$\xi_{\uH}(\alpha)$, 
		$\sqrt{c}\,\xi_{\uA}(\alpha,N)$, resp.
		$\xi_{\uF}$,
		with corresponding dimension $d=1,1,2,3$.
	\end{lem}
	
	\begin{widetext}
	Let us consider the proof for the 1D Calogero-Sutherland 
	case.
	Using the separation of the center-of-mass 
	$n\sum_j \partial_j^2 = (\sum_j \partial_j)^2 + \sum_{j<k} (\partial_j - \partial_k)^2,$
	the (Neumann) kinetic energy for $n\ge 2$ particles 
	on an interval $Q = [a,b]$ is
	\begin{multline} \label{relative_kinetic_bound}
		\int_{Q^n} \bar{\psi} \,\hat{T}_{\uH} \psi \,dx 
		\ge \int_{Q^n} \sum_{j<k} \bar{\psi} \left( -\frac{1}{2n} (\partial_j - \partial_k)^2
			+ V_{\uH}(x_j - x_k) \right) \psi \,dx \\
		\ge \frac{2}{n} \sum_{j<k}
		\int\limits_{Q^{n-2}} \int\limits_Q \int\limits_{[-\delta(R),\delta(R)]}
			\!\!\!\!\!\bar{\psi}\left( -\partial_r^2 + V_{\uH}(r) \right) \psi \,dr \,dR \,dx' \\
		\ge \frac{2}{n} \sum_{j<k}
		\int\limits_{Q^{n-2}} \int\limits_Q \frac{\xi_{\uH}(\alpha)^2}{\delta(R)^2}
			\int\limits_{[-\delta(R),\delta(R)]} |\psi|^2 \,dr \,dR \,dx',
	\end{multline}
	where for each particle pair we have introduced
	$R := (x_j+x_k)/2$, $r := x_j - x_k$, 
	$x' = (x_1,\ldots,x\!\!\!\!\diagup\!_j,\ldots,x\!\!\!\!\diagup\!_k,\ldots,x_N)$,
	and $\delta(R) := 2\min \{|R-a|,|R-b|\}$.
	We then use \eqref{relative_kinetic_bound} and
	$\delta(R)^{-2} \ge |Q|^{-2}$
	to obtain \eqref{local_exclusion_1H}.
	Inserting the partition of unity
	$1 = \sum_{A \subseteq \{1,\ldots,N\}} \prod_{l \in A} \chi_Q(x_l) \prod_{l \notin A} \chi_{Q^c}(x_l)$
	into 
	the expression for $T^Q_{\uH}$
	we then obtain
	(cp. \cite{Dyson-Lenard:67,Dyson-Lenard:reviews,Lundholm-Solovej:anyon,Lundholm-Solovej:extended})
	\begin{multline*}
		T^Q_{\uH} = \sum_A \int_{\R^N} \sum_{j \in A} \frac{1}{2} \left( 
			|\partial_j \psi|^2 + \sum_{(j\neq)k=1}^N V_{\uH}(x_j-x_k) |\psi|^2
			\right) \prod_{l \in A} \chi_Q(x_l) \prod_{l \notin A} \chi_{Q^c}(x_l) \,dx \\
		\ge \sum_A \int_{(Q^c)^{N-|A|}} \int_{Q^{|A|}} \frac{1}{2} \left(
			\sum_{j \in A} |\partial_j \psi|^2 
			+ \sum_{j \neq k \,\in A} V_{\uH}(x_j - x_k) |\psi|^2
			\right) \prod_{l \in A} dx_l \prod_{l \notin A} dx_l \\
		\ge \sum_A (|A|-1) \frac{\xi_{\uH}(\alpha)^2}{|Q|^2} 
			\int_{(Q^c)^{N-|A|}} \int_{Q^{|A|}} |\psi|^2 \prod_{l \in A} dx_l \prod_{l \notin A} dx_l \quad
		= \frac{\xi_{\uH}(\alpha)^2}{|Q|^2} \int_{\R^N} 
			\left( \sum_{j=1}^N \chi_Q(x_j) - 1 \right) |\psi|^2 dx,
	\end{multline*}
	where in the last step we again used the partition of unity.
	This proves \eqref{local_exclusion_density_all}
	in the $\alpha \ge 1$ Calogero-Sutherland 
	case.
	The Lieb-Liniger 
	case follows similarly, while 
	in the anyon case
	the application of 
	the above mentioned pairwise magnetic operator inequality gives 
	rise to a local repulsive inverse-square pair
	potential, with its strength 
	measured by \eqref{anyon_constant}. 
	We refer to \cite{Lundholm-Solovej:anyon,Lundholm-Solovej:extended} 
	for the detailed proofs.
	\end{widetext}

\begin{figure}[t]
		\centering
		\includegraphics[scale=0.34]{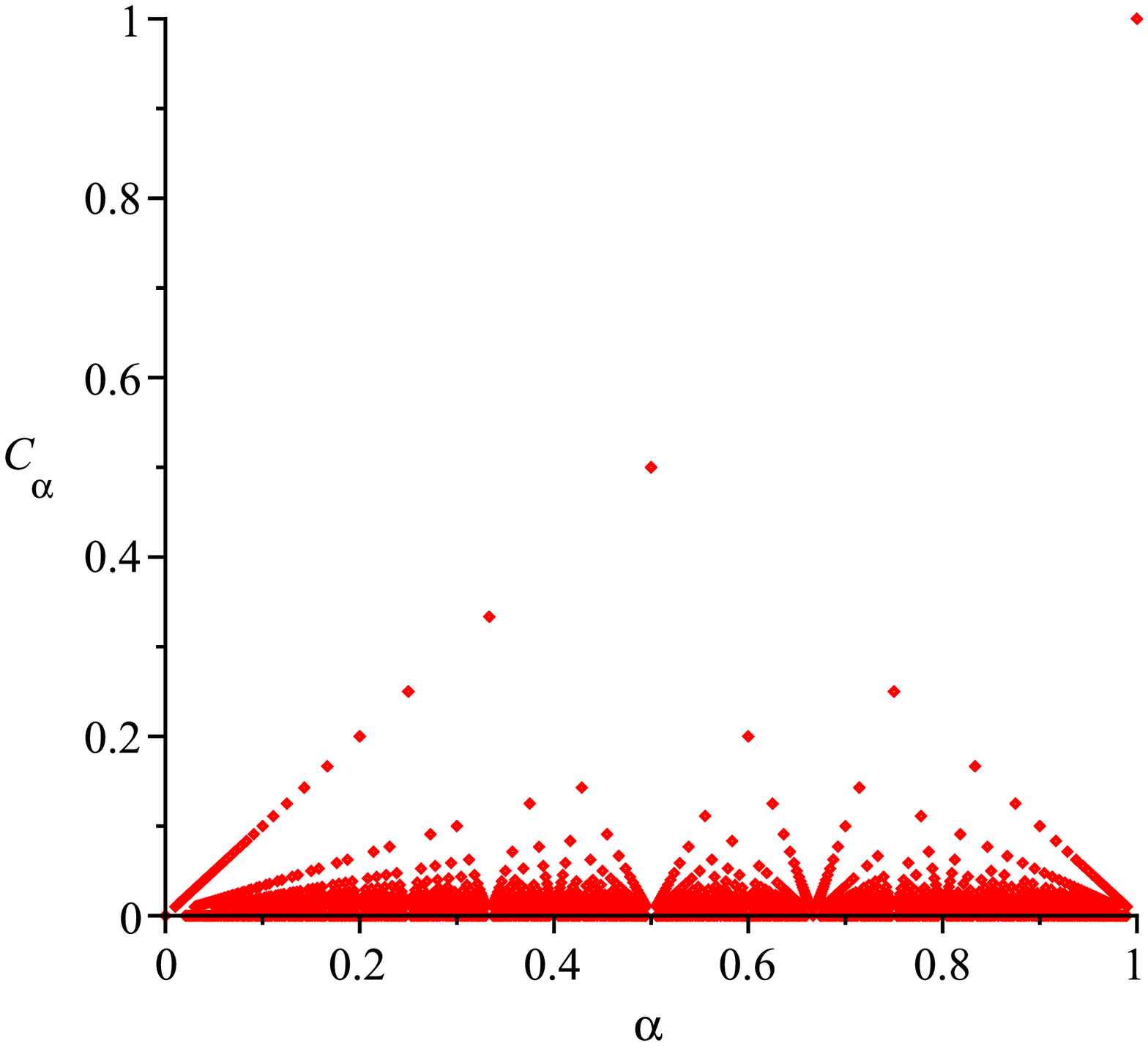}
		\caption{A sketch of 
		$C_\alpha = \lim\limits_{N \to \infty} \xi_{\uA}(\alpha,N)$ 
		as a function of $\alpha$.}
		\label{fig:xiA}
	\end{figure}

	The constants $\xi_{\uS/\uH/\uA/\uF}^2$ of proportionality in 
	\eqref{local_exclusion_density_all}
	appear as lower bounds on the strength of local exclusion,
	and could e.g. be compared with the global constant of 
	proportionality in Haldane's generalized exclusion statistics 
	\cite{Haldane:91}.
	For the case of anyons, the constant 
	$\xi_{\uA} \propto \xi_{\uA}(\alpha,N)$ 
	is actually $N$-dependent,
	and it is clear from the definition \eqref{anyon_constant} 
	that this constant can become identically zero for sufficiently
	large $N$ if $\alpha$ is an even numerator (reduced) fraction.
	However, we have shown in \cite{Lundholm-Solovej:anyon} that for
	$\alpha = \mu/\nu$ an \emph{odd numerator fraction}, the limiting
	constant is non-zero and equal to 
	$\lim_{N \to \infty} \xi_{\uA}(\alpha,N) = 1/\nu$ 
	(see Fig. \ref{fig:xiA}).
	It hence also becomes weaker with a bigger denominator $\nu$
	in the statistics parameter.
	For irrational $\alpha$ the constant is non-zero for all finite $N$,
	but the limit is again zero.
	We will return to a discussion on the true dependence on 
	$\alpha$ for the exclusion and statistics of anyons below.

\section{Lieb-Thirring-type inequalities}

	The inequalities \eqref{LT} and \eqref{kinetic-LT} for fermions 
	combine the Pauli exclusion principle with the uncertainty principle
	to produce non-trivial and useful bounds for the energy 
	as the number of particles $N$ becomes large.
	We shall complement the local form of the exclusion principle above
	with the following 
	\emph{local form of the uncertainty principle}
	on a $d$-dimensional cube $Q$,
	valid for the free kinetic energy of any bosonic wave function $\psi$,
	and hence applicable in our cases of intermediate statistics 
	after discarding the positive statistics potentials
	or, in the case of anyons, using the diamagnetic inequality 
	$|D_j \psi| \ge |\nabla_j |\psi||$:
	\begin{equation} \label{local_uncertainty}
		T_{0/\uS/\uH/\uA}^Q \ \ge \ 
		c_1 \frac{ \int_Q \rho^{1+2/d} \,d\bx }{( \int_Q \rho \,d\bx )^{2/d}}
			- c_2 \frac{ \int_Q \rho \,d\bx }{|Q|^{2/d}}.
	\end{equation}
	The constants $c_2 > c_1 > 0$ only depend on $d$.
	Mathematically, \eqref{local_uncertainty} 
	is a form of Poincar\'e-Sobolev inequality, and we refer to 
	\cite{Lundholm-Solovej:anyon,Lundholm-Solovej:extended,Lundholm-Portmann-Solovej} 
	for details and proofs.
	Note that the r.h.s. is bigger for less constant density,
	but scales with the number of particles only as $N$
	(in contrast to the Lieb-Thirring inequality).

	Combining local uncertainty with local exclusion, 
	and cleverly splitting the space into smaller cubes 
	depending on the density
	(the bound \eqref{local_exclusion_density_all} 
	is strongest for cubes $Q$ s.t. 
	the expected number of particles to be found on $Q$ is
	$\int_Q \rho \approx 2$),
	one can then prove the following energy bounds:

	\begin{thm}[L-T inequalities for anyons]
		For any $\alpha \in \R$, 
		the free kinetic energy for $N$ anyons satisfies the bound
		\begin{equation} \label{kinetic-LT-anyon}
			T_{\uA} 
			\ \ge \ C_{\uA} \, \xi_{\uA}(\alpha,N)^2 
				\int_{\R^2} \rho(\bx)^2 \,d\bx,
		\end{equation}
		for some constant 
		$10^{-4} \le C_{\uA} \le \pi$.
		It follows that
		if $\alpha = \mu/\nu$ is 
		a reduced fraction with odd numerator $\mu$
		and the density $\rho$ is supported on an area $L^2$ then
		\begin{equation} \label{energy-anyon-gas}
			T_{\uA}/L^2 \ \ge \ C_{\uA} \frac{\bar{\rho}^2}{\nu^2},
			\qquad \bar{\rho} := N/L^2.
		\end{equation}
	\end{thm}

	\begin{thm}[L-T inequalities for 1D Lieb-Liniger] 
		For $\eta \ge 0$
		\begin{equation} \label{kinetic-LT-Schrodinger}
			T_{\uS} 
			\ \ge \ C_{\uS} \int_{\R} \xi_{\uS}(2\eta/\rho(x))^2 \rho(x)^3 \,dx,
		\end{equation}
		for some constant 
		$10^{-5} \le C_{\uS} \le 2/3$. 
		In particular, if $\rho$ is homogeneous,
		e.g. $\rho \le \gamma \bar{\rho}$ for some $\gamma>0$, then
		\begin{equation} \label{kinetic-LT-SchrodingerBd}
			T_{\uS} 
			\ \ge \ C_{\uS} \,\xi_{\uS}(2\eta/(\gamma\bar{\rho}))^2 \int_{\R} \rho(x)^3 \,dx,
		\end{equation}
		and if $\rho$ is supported on an interval of length $L$
		\begin{equation} \label{energy-SchrodingerBd-gas}
			T_{\uS}/L \ \ge \ C_{\uS} \,\xi_{\uS}(2\eta/(\gamma\bar{\rho}))^2 \bar{\rho}^3,
			\qquad \bar{\rho} := N/L.
		\end{equation}
	\end{thm}

	It is illustrative to
	compare with Lieb and Liniger \cite{Lieb-Liniger:63}, where 
	for a free system in the thermodynamic limit 
	$N,L \to \infty$ with fixed density $\bar{\rho}$,
	$T_{\uS}/L \to \frac{1}{2} e(2\eta/\bar{\rho}) \bar{\rho}^3$ 
	with $e(t) \sim t, t \ll 1$, $e(t) \to \frac{\pi^2}{3}, t \to \infty$
	(see also \cite{Lieb-Seiringer-Yngvason:03}). 

	\begin{thm}[L-T inequalities for 1D C.-S.] 
		For $\alpha \ge 1$ and arbitrary intervals $Q$
		such that the expected number of particles
		$\int_Q \rho \ge 2$
		\begin{equation} \label{kinetic-LT-Heisenberg-local}
			T_{\uH}^Q \ \ge \ C_{\uH} \, \xi_{\uH}(\alpha)^2 \frac{\left( \int_Q \rho (x) \,dx \right)^3}{|Q|^2},
		\end{equation}
		with a constant 
		$1/32 \le C_{\uH} \le 2/3$.
		It follows in particular that if $\rho$ is 
		confined to a length $L$ and $N \ge 2$ then
		\begin{equation} \label{energy-Heisenberg-gas}
			T_{\uH}/L \ \ge \ C_{\uH} \,\xi_{\uH}(\alpha)^2 \bar{\rho}^3,
			\qquad \bar{\rho} := N/L.
		\end{equation}
	\end{thm}

	Compare with Calogero and Sutherland 
	\cite{Calogero:69 & Sutherland:71}, where one finds
	$T_{\uH}/L \to \frac{\pi^2}{6} \alpha^2 \bar{\rho}^3$
	in the thermodynamic limit $N,L \to \infty$.

	The reason for the more technical forms 
	\eqref{kinetic-LT-Schrodinger} and \eqref{kinetic-LT-Heisenberg-local}
	as compared to \eqref{kinetic-LT} and \eqref{kinetic-LT-anyon}
	is the local dependence of the strength of exclusion in the Lieb-Liniger 
	case, respectively the possibility for arbitrarily strong exclusion
	($\alpha \to \infty$) in the Calogero-Sutherland 
	case.
	We sketch a proof below only for the simpler anyonic case, 
	and refer to \cite{Lundholm-Solovej:anyon,Lundholm-Solovej:extended,Lundholm-Portmann-Solovej} 
	for further details. 
	For an application of the same method to fermions in 3D
	and the generalization to $q$ spin states
	we refer to \cite{Frank-Seiringer:12}
	where a model for point interactions was considered. 

\begin{figure}[t]
		\centering
		\includegraphics[scale=0.71]{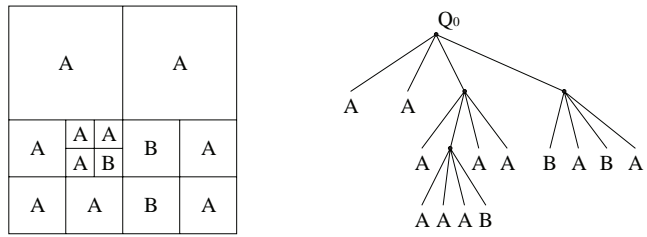}
		\caption{Example of a splitting of $Q_0$ and a corresponding
			tree $\mathbb{T}$ of subsquares. For the B-square at
			level 3 in the tree, the set $\mathcal{A}(Q)$ of all associated 
			A-squares (cp. $\mathcal{A}_1(Q)$ in the text)
			consists of 8 elements, while for the two
			B-squares at level 2, $\mathcal{A}(Q)$ coincide and
			has 4 elements.}
		\label{fig:splitting}
	\end{figure}

		Let us for simplicity
		assume $\rho$ to be supported on some square $Q_0$ in the plane
		which we proceed to split into four smaller squares iteratively, 
		organizing the resulting subsquares $Q$ in a tree $\mathbb{T}$ 
		(see Fig. \ref{fig:splitting}).
		The procedure can be arranged so that $Q_0$
		is finally covered by a collection $Q \in \mathbb{T}_B$
		of non-overlapping squares 
		marked B s.t. 
		$2 \le \int_{Q} \rho < 8$, and $Q \in \mathbb{T}_A$ 
		marked A s.t. $0 \le \int_{Q} \rho < 2$,
		and s.t. at least one B-square is at the topmost level 
		of every branch of the tree $\mathbb{T}$.
		On the B-squares we use 
		\eqref{local_exclusion_density_all} together with 
		\eqref{local_uncertainty} 
		to obtain
		(with $c_k'>0$ some numerical constants)
		\begin{equation} \label{B-squares}
			T_{\uA}^{Q} \ge \xi_{\uA}(\alpha,N)^2
			\left( c_1' \int_{Q} \rho^2 + \frac{c_2'}{|Q|} \right), 
			\ \ Q \in \mathbb{T}_B.
		\end{equation}
		The A-squares are further grouped into a subclass A$_2$ 
		on which the density is sufficiently non-constant,
		$
			\int_{Q} \rho^2 > \frac{2c_2}{c_1} (\int_{Q} \rho)^2/|Q|$ for $Q \in \mathbb{T}_{A_2} \subseteq \mathbb{T}_A
		$,
		so that by 
		\eqref{local_uncertainty} 
		\begin{equation}
			T_{\uA}^{Q} > \frac{c_1}{4} \int_{Q} \rho^2,
			\quad Q \in \mathbb{T}_{A_2},
		\end{equation}
		and a subclass A$_1$ on which
		$
			\int_{Q} \rho^2 \le \frac{2c_2}{c_1} (\int_{Q} \rho)^2/|Q|
		$.
		One can then use the structure of the splitting of squares 
		to prove that, for the set $\mathcal{A}_1(Q_B)$
		of all such A$_1$-squares
		which can be found by stepping backwards in the tree $\mathbb{T}$ 
		from a fixed B-square $Q_B$ 
		and then one step forward,
		\begin{equation}
			\sum_{Q \in \mathcal{A}_1(Q_B)} \int_{Q} \rho^2 
			\le \sum_{k=0}^\infty 3 \frac{2c_2}{c_1} \frac{2^2}{4^{k}|Q_B|}
			= \frac{32c_2}{c_1} \frac{1}{|Q_B|}.
		\end{equation}
		In other words the energy on all subsquares with constant low
		density is dominated by that from exclusion on the B-squares.
		We therefore find from \eqref{B-squares}
		that 
		$$
			T_{\uA}^{Q_B} \ge \xi_{\uA}(\alpha,N)^2 \left(
			c_1' \int_{Q_B} \rho^2 
			+ c_3' \sum_{Q \in \mathcal{A}_1(Q_B)} 
				\int_Q \rho^2 \right),
		$$
		and hence,
		since all A$_1$-squares are covered in this way,
		$
			T_{\uA} = \sum_{Q \in \mathbb{T}} T_{\uA}^{Q} 
			\ge \sum_{Q \in \mathbb{T}_B \cup \mathbb{T}_{A_2}} T_{\uA}^{Q} 
			\ge C_{\uA} \,\xi_{\uA}(\alpha,N)^2 \int_{Q_0} \rho^2
		$
		for some numerical constant $C_{\uA} >0$.

	For \eqref{energy-SchrodingerBd-gas} 
	and \eqref{energy-anyon-gas} we use 
	$\int_{Q_0} \rho^p dx \ge N^p |Q_0|^{1-p}$,
	and for \eqref{energy-anyon-gas} we used the fact that
	$\lim_{N \to \infty} \xi_{\uA}(\alpha,N) = 1/\nu$ for such 
	odd numerator fractions and zero otherwise.

\section{An application to harmonic oscillator confinement}

	Consider $N$ anyons with statistics parameter $\alpha$
	confined in an external one-body harmonic oscillator 
	potential $V(\bx) = \frac{\omega^2}{2}|\bx|^2$.
	Using the bound \eqref{kinetic-LT-anyon} for the kinetic
	energy we obtain as a lower bound for the total energy the
	following functional of the density:
	\begin{equation} \label{harmonic-oscillator-functional}
		T_{\uA} + \langle \hat{V} \rangle_{\psi}
			\ge \int_{\R^2} \left( 
				C_{\uA} \xi_{\uA}(\alpha,N)^2 \rho(\bx)^2 
				+ \frac{\omega^2}{2} |\bx|^2 \rho(\bx) \right) d\bx.
	\end{equation}
	It is straightforward \cite{Lundholm-Solovej:extended}
	to extremize this functional w.r.t. $\rho$
	under the constraint $\int_{\R^2} \rho = N$
	to obtain the minimizer
	$$
		\rho(\bx) = 
		\frac{\left( \omega \xi_{\uA}(\alpha,N) \sqrt{2C_{\uA}N/\pi} 
			- \omega^2|\bx|^2/2 \right)_+}{2C_{\uA}\xi_{\uA}(\alpha,N)^2},
	$$
	and therefore the (rigorous) lower bound for 
	the ground state energy $E_0$:
	\begin{equation} \label{harmonic-oscillator-bound}
		T_{\uA} + \langle \hat{V} \rangle_{\psi} 
		\ \ge \ E_0 \ \ge \ 
		\frac{1}{3}\sqrt{\frac{8C_{\uA}}{\pi}} \,\xi_{\uA}(\alpha,N) \,\omega N^{3/2}.
	\end{equation}
	In the case of 
	odd numerator rational $\alpha$ this improves the bound 
	given in \cite{Chitra-Sen:92}
	(which is also valid for arbitrary $\alpha$):
	\begin{equation} \label{anyon-momentum-bound}
		E_0
		\ \ge \ \omega \left( N + \left|L + \alpha \frac{N(N-1)}{2} \right| \right),
	\end{equation}
	where $L$ denotes the total angular momentum of the state $\psi$.
	Note that if $L = -\alpha \binom{N}{2}$
	(which could occur for certain $N$ and rational 
	$\alpha$ as long as the r.h.s. is an even integer)
	then this bound reduces to the bosonic bound for the energy,
	which is always valid as a trivial lower bound.

	It was in \cite{Chitra-Sen:92} argued using perturbation theory 
	that the behavior for the exact ground state energy 
	as $N \to \infty$ is approximately 
	$E_0 \sim \sqrt{\alpha}\omega N^{3/2}$ for $\alpha \sim 0$
	and $E_0 \sim \frac{1}{3}\sqrt{8}\omega N^{3/2}$ for $\alpha \sim 1$, 
	requiring $L = -\alpha \binom{N}{2} + O(N^{3/2})$
	by \eqref{anyon-momentum-bound}.
	In fact, we can show that the ground state energy necessarily always
	satisfies the upper bound $E_0 \lesssim \omega N^{3/2}$, 
	up to a constant independent of $\alpha$. 
	Namely, given a (possibly non-symmetric) $N$-particle wave function 
	$\phi$ s.t. all particles are supported on disjoint sets,
	we can form its symmetrization
	\begin{equation} \label{symmetrization}
		\psi(x) := \frac{1}{\sqrt{N!}} \sum_{\sigma \in S_N} \phi(\bx_{\sigma(1)},\ldots,\bx_{\sigma(N)})
	\end{equation}
	and conclude by the properties of the supports 
	that $\|\psi\| = \|\phi\|$ 
	and $\int \sum_j |D_j \psi|^2 dx = \int \sum_j |D_j \phi|^2 dx$.
	Now, take e.g. 
	\begin{equation} \label{gauge-support-phi}
		\phi(x) := \prod_{j=1}^N \varphi(\bx_j - \by_j) \prod_{k<l} e^{-i\alpha \phi_{kl}},
	\end{equation}
	where $\by_j$ are fixed points in the plane 
	separated by a minimal length $r$,
	the function $\varphi$ localizes each particle in a ball of radius $r/2$,
	and $\phi_{kl}$ is the angle between particle $k$ and $l$ relative to
	a fixed axis. 
	Note that this angle is well defined and smooth,
	and that the resulting phase factor (gauge) cancels the magnetic potentials 
	$\bA_j$ in $D_j \phi$.
	Then, by choosing the points $\by_j$ to fill a disk of radius 
	$R \sim \sqrt{N} r$,
	we conclude that the energy $E$ of $\psi$ is bounded by
	$E \lesssim N/r^2 + \omega^2 NR^2 \sim N^2/R^2 + \omega^2 NR^2$,
	and hence, choosing $r$ s.t. $R^2 \sim \sqrt{N}/\omega$, 
	we have $E \lesssim \omega N^{3/2}$.
	It therefore follows that, for odd numerator $\alpha$, \eqref{harmonic-oscillator-bound} 
	yields the correct dependence in $N$ up to the
	value of the constant.
	In a similar way one can also prove that the ground state energy
	per unit area
	for the free anyon gas is uniformly bounded by a constant times 
	$\bar{\rho}^2$.

	For comparison, we can also consider the 1D Calogero-Sutherland case
	together with an external potential
	--- for which some exact results are available \cite{Calogero:69 & Sutherland:71}.
	After splitting the real line into intervals big enough to contain 
	a sufficient number of particles,
	the local bound \eqref{kinetic-LT-Heisenberg-local}
	can be applied 
	on each such interval
	with the addition of an external potential, thereby obtaining 
	an energy functional and lower estimates for the ground state energy 
	(along with estimates for the corresponding ground state density). 
	Depending on the potential, the lower bound can be 
	optimized to the better
	by choosing the splitting suitably.
	As an example, we considered in \cite{Lundholm-Solovej:extended}
	the external one-body potential $V(x) = a^\mu |x|^\mu$
	and obtained a lower bound for the total ground state energy
	\begin{equation} \label{Heisenberg-potential-bound}
		T_{\uH} + \langle \hat{V} \rangle_{\psi} 
		\ \ge \ E_0 \ \ge \ 
		C(\mu) (\xi_{\uH}(\alpha) a)^{\frac{2\mu}{\mu+2}} N^{\frac{3\mu+2}{\mu+2}},
	\end{equation}
	in the limit $N \to \infty$ and
	with an explicit constant $C(\mu)$.
	In the harmonic oscillator case $\mu=2$, $a=\omega/\sqrt{2}$,
	one obtains $E_0 \ge \frac{\sqrt{3}}{8\pi} \xi_{\uH}(\alpha) \omega N^2$,
	which can be compared to the exact ground state energy for 
	the Calogero-Sutherland model,
	$E_0 = \frac{1}{2} \omega N(1 + \alpha(N-1))$.
	We can also compare these rigorous bounds with the approximate 
	Thomas-Fermi theory \cite{Sen-Bhaduri:95 & Smerzi:96}
	and collective field theory \cite{Sen-Bhaduri:97}.

\section{Discussion}

	The bound \eqref{energy-anyon-gas} provides a non-trivial lower
	bound for the energy per unit area for an ideal gas
	of anyons with odd-fractional statistics parameter $\alpha$.
	The numerical constant $C_{\uA} \ge 10^{-4}$ in this bound has in 
	\cite{Lundholm-Solovej:extended} 
	been improved to $\ge 0.021$,
	which is still quite far from the exact semiclassical constant 
	$\pi$ for the two-dimensional spinless fermion gas.
	In any case, 
	these non-trivial bounds \eqref{energy-anyon-gas} 
	and \eqref{harmonic-oscillator-bound} raise
	the very interesting question of whether such Lieb-Thirring inequalities
	are in fact not valid 
	for even numerator and irrational $\alpha$.
	We give some motivation for why this could be the case by considering
	the following observations.

	In these bounds the expression $\xi_{\uA}(\alpha,N)$
	appears as a measure of exclusion. Its complicated behavior
	in $\alpha$ and $N$ 
	is related to the fact that only for even numerator 
	fractions do the anyons appear to have the possibility to completely 
	cancel out the statistical phase which is responsible for a 
	local repulsive force between them,
	by assuming certain configurations. 
	Consider e.g. a pair of $\alpha=2/3$ anyons which symmetrically encircle 
	a third one with relative angular momentum $-2$, leading to a local 
	cancellation of the interchange phase with the orbital phase, 
	and $\xi_{\uA}(\alpha,3) = 0$. 
	A similar complete cancellation would never be possible for odd 
	numerator $\alpha$, and indeed $\xi_{\uA}(\alpha,N)$ 
	remains strictly bounded away from zero
	for all $N$.

	Let us again consider the model of $N$ anyons in a harmonic 
	oscillator potential.
	It is well-known that energy levels and degeneracies 
	in this model depend 
	very non-trivially on both $N$ and $\alpha$,
	and we can 
	point out certain similarities 
	in the limiting graph of $\xi_{\uA}(\alpha,N)$ 
	(see Fig. \ref{fig:xiA})
	with known features in spectra for $N=2,3,4$ 
	and corresponding extrapolations to large $N$ 
	\cite{Leinaas-Myrheim:77,Chitra-Sen:92,Sporre-Verbaarschot-Zahed:91-92 & Sen:92,Khare:05}.
	It is e.g. intriguing to compare 
	this graph
	--- which can be obtained by cutting out a wedge of slope $\nu$
	from the upper half-plane 
	at every even numerator rational point $\mu/\nu$ on the horizontal axis
	--- with the general structure indicated in Fig. 1 in \cite{Chitra-Sen:92}.

	The question remains whether for particular $\alpha$
	(even numerator rational and/or irrational)
	the energy could be of 
	lower order than $O(N^{3/2})$
	for some special states with $L \sim -\alpha\binom{N}{2}$.
	With the above considerations, and
	motivated by the Laughlin states in the 
	fractional quantum Hall effect
	\cite{Laughlin:83},
	we could for particular $N=\nu K$ consider 
	trial wave functions of the form
	$\psi = \Phi \psi_\alpha$, with
	\begin{equation} \label{trial-even}
		\psi_\alpha := \prod_{j<k} |z_{jk}|^{-\alpha} 
			\,\Sc\left[ \prod_{q=1}^\nu 
			\prod_{(j,k) \in \mathcal{E}_q} (\bar{z}_{jk})^\mu 
			\prod_{l \in \mathcal{V}_q} \varphi_0(\bx_l) 
			\right]
	\end{equation}
	for even numerator fractions $\alpha = \mu/\nu \in [0,1]$, 
	and 
	\begin{equation} \label{trial-odd}
		\psi_\alpha := \prod_{j<k} |z_{jk}|^{-\alpha} 
			\,\Sc\left[ \prod_{q=1}^\nu 
			\prod_{(j,k) \in \mathcal{E}_q} (\bar{z}_{jk})^\mu 
			\bigwedge_{k=0}^{K-1} \varphi_k \,(\bx_{l \in \mathcal{V}_q})
			\right]
	\end{equation}
	for odd numerators $\mu$, 
	where the role of the factor
	$\Phi$ is to 
	regularize the short-scale behavior
	(necessary due to the singular Jastrow factor
	in \eqref{trial-even} and \eqref{trial-odd}).
	We have written 
	$z_{jk}:=z_j - z_k$ for the complex relative particle coordinates, 
	$\varphi_k$ denote the eigenstates of the one-particle Hamiltonian
	$\hat{h} = -\frac{1}{2}\nabla^2 + V$
	and of which we may form a Slater determinant $\bigwedge_k \varphi_k$, 
	while $\mathcal{E}_q$ and $\mathcal{V}_q$ 
	are sets of edges and vertices 
	of $\nu$ disjoint
	complete graphs involving $K$ particles each,
	and $\Sc$ denotes the operation of symmetrization 
	(cp. \eqref{symmetrization}).
	Two possible choices of regularizing symmetric functions $\Phi$, 
	giving rise to the expected pairwise short-scale behavior 
	$\sim |z_{jk}|^\alpha$ in $\psi$,
	could be
	\begin{equation} \label{Phi-parameter}
		\Phi_{r_0} = \prod_{j<k}|z_{jk}|^{2\alpha}(r_0^2 + |z_{jk}|^2)^{-\alpha},
	\end{equation}
	with a parameter $r_0>0$, or the parameter-free (but less smooth)
	\begin{equation} \label{Phi-neighbor}
		\Phi = \prod_{j=1}^N \prod_{k=1}^{\nu-1} |z_{j\,k(j)}|^{\alpha},
	\end{equation}
	with $k(j)$ denoting the $k$th nearest neighbor of particle $j$.
	These states $\psi$ have 
	$L = -\alpha\binom{N}{2} + \alpha\frac{\nu-1}{2}N$
	(for \eqref{trial-even} and for certain magic numbers 
	$K$ in \eqref{trial-odd})
	and the property that only up to $\nu$ particles can 
	be selected in each term
	without involving a repulsive factor $(\bar{z}_{jk})^\mu$ from 
	an edge in $\mathcal{E}_q$ for some $q$,
	allowing for the formation of groups of $\nu$ 
	anyons with integer statistics flux $\mu$.
	Namely, while the Jastrow factor acts to attract all particles,
	this attraction is on large scales 
	exactly balanced whenever a group of $\nu$ non-repelling anyons
	has formed, since an anyon $\bx_j$ far outside the group, 
	seeing the total attractive factor $\sim (r^{-\alpha})^\nu = r^{-\mu}$ 
	where $r$ is the distance from the group,
	is also repelled by at least one anyon $\bx_k$ in the group, 
	with a factor $|\bar{z}_{jk}|^\mu \sim r^\mu$ 
	from that corresponding edge in $\mathcal{E}_q$.
	This balance could act to distribute the anyons, on the average, 
	in such groups of $\nu$.
	Furthermore, the total contribution from such a group 
	to the statistics potential $\bA_j$ seen by the distant particle $\bx_j$ 
	would be $\sim \nu \alpha \br I/r^2 = \mu \br I/r^2$, 
	while the particle also has an orbital angular momentum $-\mu$
	around the group 
	(due to that same edge to $\bx_k$ and phase of $(\bar{z}_{jk})^\mu$)
	with velocity $\sim - \mu \br I/r^2$,
	again leading to a 
	cancellation of terms in the kinetic energy $D_j\psi$.
	
	The forms \eqref{trial-even} and \eqref{trial-odd} bring out 
	a structural difference between even and odd numerators $\mu$.
	The limit $\alpha=1$ of \eqref{trial-odd} is the fermionic 
	ground state in the bosonic representation, 
	and also generalizes to the correct gauge copies for 
	arbitrary integer $\alpha$,
	while the states with $\nu>1$ in \eqref{trial-even} are 
	(modulo the Jastrow factor) 
	actually found to be exactly the Read-Rezayi states 
	for fractional quantum Hall liquids in their bosonic form
	\cite{Read-Rezayi-states}.
	The state \eqref{trial-even} is an exact but singular 
	(requiring the regularization by $\Phi$)
	eigenstate of the Hamiltonian with energy 
	$E = \omega(N + \deg \psi_\alpha)$, 
	where $\deg \psi_\alpha = -\alpha\frac{\nu-1}{2}N$  
	is the total degree of the non-Gaussian part of the wave function
	(cp. \cite{Chou:91 & Bhaduri_et_al:92}).
	In all known exact eigenstates there is this simple correspondence between
	the degree and the energy. It is an interesting fact that
	adding the degree of $\Phi$ in the nearest-neighbor form 
	\eqref{Phi-neighbor}
	produces $\omega(1 + \alpha\frac{\nu-1}{2})N$, i.e. exactly 
	the r.h.s. of \eqref{anyon-momentum-bound}
	for the above value of $L$,
	speaking for a low energy for even numerator fractions.
 	On the other hand, the degree of the odd numerator states
 	\eqref{trial-odd} necessarily grows with $K$ as $\sim K^{3/2}$
 	due to the Slater determinant.
 	While the resulting energy
	$E = \omega(N + \deg \psi_\alpha) \sim \omega \nu (N/\nu)^{3/2}$
	satisfies but does not match the bound \eqref{harmonic-oscillator-bound}
	exactly w.r.t. $\alpha$,
	a corresponding picture of ideal anyons forming essentially free 
	$\nu$-anyon groups
	with fermionic type statistics would actually match the form of the bound
	\eqref{energy-anyon-gas}, involving the reduced density 
	$\bar{\rho}/\nu = K/L^2$.
	
	We finally remark that
	there are also many interesting connections between the forms
	of the fractions appearing here and those of fractionally charged
	quantum Hall quasiparticles
	\cite{Lee-Fisher:89 & Moore-Read:91}.
	Another question concerns possible relations with 
	$q$-commutation relations, with $q=e^{i\alpha\pi}$ 
	\cite{Goldin-Sharp:96 & Goldin-Majid:04}.

	\begin{acknowledgments}
		Support from the Danish Council for Independent Research 
		as well as from Institut Mittag-Leffler (Djursholm, Sweden)
		is gratefully acknowledged. 
		D.L. also thanks
		IH\'ES, IHP, FIM ETH Zurich, and the Isaac Newton Institute 
		(EPSRC Grant EP/F005431/1)
		for support and hospitality via 
		EPDI and CARMIN fellowships.
		J.P.S. acknowledges support by ERC AdGrant Project No. 321029.
		This work was partly done
		while participating in the research programs 
		``Hamiltonians in Magnetic Fields'' 
		at Institut Mittag-Leffler and 
		``Variational and Spectral Methods in Quantum Mechanics''
		at Institut Henri Poincar\'e. 
		We thank E. Ardonne, J. Derezinski, G. Felder, J. Fr\"ohlich, 
		G. Goldin, H. Hansson, J. Hoppe, T. Jolicoeur, E. Langmann, 
		S. Ouvry, F. Portmann, 
		N. Rougerie, R. Seiringer and J. Yngvason 
		for comments and discussions.
	\end{acknowledgments}


\begin{thebibliography}{99}

\bibitem{Streater-Wilde:70}
R. F. Streater, I. F. Wilde, 
Nucl. Phys. B 24, 561 (1970). 

\bibitem{Leinaas-Myrheim:77}
J. M. Leinaas, J. Myrheim, 
Il Nuovo Cimento 37B, 1 (1977). 

\bibitem{Goldin-Menikoff-Sharp:81 & Wilczek:82}
G. A. Goldin, R. Menikoff, D. H. Sharp, 
J. Math. Phys. 22, 1664 (1981); 
F. Wilczek, 
Phys. Rev. Lett 48, 1144 (1982); 
\textit{ibid.} 49, 957 (1982). 

\bibitem{ref:trapped-bosons}
T. Kinoshita, T. Wenger, D. S. Weiss, Science 305, 1125 (2004);
B. Paredes \textit{et al.}, Nature 429, 277 (2004).

\bibitem{ref:QHE}
See e.g. R. B. Laughlin, Rev. Mod. Phys. 71, 863 (1999);
R. E. Prange, S. M. Girvin (eds.), \textit{The Quantum Hall Effect} (Springer-Verlag, Second Edition 1990).

\bibitem{Myrheim:99}
J. Myrheim, \textit{Anyons}, in Topological aspects of low dimensional systems (Les Houches, 1998), pp. 265--413, EDP Sci., Les Ulis, 1999.

\bibitem{Khare:05}
A. Khare, \textit{Fractional Statistics and Quantum Theory}, (World Scientific, Singapore, Second Edition 2005).

\bibitem{ref:reviews}
I. Bloch, J. Dalibard, W. Zwerger, Rev. Mod. Phys. 80, 885 (2008);
J. Fr\"ohlich, \textit{Quantum statistics and locality}, in Proceedings of the Gibbs Symposium (New Haven, CT, 1989), pp. 89--142, Amer. Math. Soc., Providence, RI, 1990;
A. Lerda, \textit{Anyons}, (Springer-Verlag, Berlin--Heidelberg, 1992);
S. Ouvry, S\'em. Poincar\'e XI, 77 (2007);
F. Wilczek, \textit{Fractional Statistics and Anyon Superconductivity}, (World Scientific, Singapore, 1990).

\bibitem{Haldane:91}
F. D. M. Haldane, 
Phys. Rev. Lett. 67, 937 (1991). 

\bibitem{Isakov:94}
S. B. Isakov, 
Phys. Rev. Lett. 73, 2150 (1994). 

\bibitem{Lieb-Thirring:75}
E. H. Lieb, W. Thirring, 
Phys. Rev. Lett. 35, 687 (1975); 
\textit{Inequalities for the Moments of the Eigenvalues of the Schr\"odinger Hamiltonian and Their Relation to Sobolev Inequalities}, in Stud. Math. Phys., pp. 269--303, Princeton University Press, 1976;
see also
E. H. Lieb, R. Seiringer, \textit{The stability of matter in quantum mechanics}, (Cambridge University Press, Cambridge, 2010).

\bibitem{Dolbeault-Laptev-Loss:08}
J. Dolbeault, A. Laptev, M. Loss, J. Eur. Math. Soc. 10, 1121 (2008).

\bibitem{Gentile:40-42}
G. Gentile, Il Nuovo Cimento 17, 493 (1940); 
\textit{ibid.} 19, 109 (1942). 

\bibitem{Dyson-Lenard:67}
F. J. Dyson, A. Lenard, 
J. Math. Phys. 8, 423 (1967). 

\bibitem{Dyson-Lenard:reviews}
F. J. Dyson, \textit{Stability of Matter}, in Statistical Physics, Phase Transitions and Superfluidity, Brandeis University Summer Institute in Theoretical Physics 1966, pp. 179--239, (Gordon and Breach Publishers, New York, 1968);
A. Lenard, \textit{Lectures on the Coulomb Stability Problem}, in Statistical mechanics and mathematical problems, Battelle Rencontres, Seattle, Wash., 1971, Lect. Notes Phys., Vol. 20, pp. 114--135, 1973.

\bibitem{Lundholm-Solovej:anyon}
D. Lundholm, J. P. Solovej, 
Commun. Math. Phys. 322, 883 (2013), {\tt DOI:10.1007/s00220-013-1748-4}

\bibitem{Lundholm-Solovej:extended}
D. Lundholm, J. P. Solovej, 
Ann. Henri Poincar\'e, 2013, {\tt DOI:10.1007/s00023-013-0273-5}

\bibitem{1D-refs}
J. M. Leinaas, J. Myrheim, Phys. Rev. B 37, 9286 (1988); Int. J. Mod. Phys. A 8, 3649 (1993); 
A. P. Polychronakos, Nucl. Phys. B 324, 597 (1989); 
C. Aneziris, A. P. Balachandran, D. Sen, Int. J. Mod. Phys. A 6, 4721 (1991); 
S. B. Isakov, Mod. Phys. Lett. A 7, 3045 (1992). 

%
%
%

\bibitem{Note:defs-SH}
As emphasized in \cite{Myrheim:99},
	this definition is the natural one for all $N$ in the 
	Schr\"odinger-type approach to quantization, 
	however, the Heisenberg-type approach naturally arrives at this model 
	only for $N=2$, but is extended in this way to all $N$.
	The resulting model is relevant for anyons in the lowest Landau
	level; see e.g. 
	T. H. Hansson, J. M. Leinaas, J. Myrheim, Nucl. Phys. B 384, 559 (1992),
	and 
	S. Ouvry, Phys. Lett. B 510, 335 (2001).

\bibitem{Lieb-Liniger:63}
E. H. Lieb, W. Liniger, 
Phys. Rev. 130, 1605 (1963). 

\bibitem{Calogero:69 & Sutherland:71}
F. Calogero, 
J. Math. Phys. 10, 2197 (1969); 
B. Sutherland, 
J. Math. Phys. 12, 246 (1971). 

\bibitem{Lundholm-Portmann-Solovej}
D. Lundholm, F. Portmann, J. P. Solovej, 
{\tt arXiv:1402.4463}

\bibitem{Lieb-Seiringer-Yngvason:03}
E. H. Lieb, R. Seiringer, J. Yngvason, 
Phys. Rev. Lett. 91, 150401 (2003). 

\bibitem{Frank-Seiringer:12}
R. L. Frank, R. Seiringer, 
J. Math. Phys. 53, 095201 (2012).

\bibitem{Chitra-Sen:92}
R. Chitra, D. Sen, 
Phys. Rev. B 46, 10923 (1992). 

\bibitem{Sen-Bhaduri:95 & Smerzi:96}
D. Sen, R. K. Bhaduri, 
Phys. Rev. Lett. 74, 3912 (1995); 
A. Smerzi, 
\textit{ibid.} 76, 559 (1996). 

\bibitem{Sen-Bhaduri:97}
D. Sen, R. K. Bhaduri, 
Ann. Phys. 260, 203 (1997). 

\bibitem{Sporre-Verbaarschot-Zahed:91-92 & Sen:92}
M. Sporre, J. J. M. Verbaarschot, I. Zahed, 
Phys. Rev. Lett. 67, 1813 (1991); 
M. V. N. Murthy \textit{et al.}, 
\textit{ibid.} 67, 1817 (1991);
M. Sporre, J. J. M. Verbaarschot, I. Zahed, 
Phys. Rev. B. 46, 5738 (1992); 
D. Sen, 
Phys. Rev. D 46, 1846 (1992). 

\bibitem{Laughlin:83}
R. B. Laughlin, 
Phys. Rev. Lett. 50, 1395 (1983). 

\bibitem{Read-Rezayi-states}
See N. Read, E. Rezayi, Phys. Rev. B 59, 8084 (1999),
and the bosonic version given in A. Cappelli \textit{et al.}, Nucl. Phys. B 599, 499 (2001).
We were not aware of this interesting coincidence at the time 
we first discovered these anyonic trial states.
It is in this context amusing to speculate whether non-abelian anyons
could arise as quasiparticle excitations of such abelian anyon states.

\bibitem{Chou:91 & Bhaduri_et_al:92}
C. Chou, 
Phys. Lett. A 155, 245 (1991); 
R. K. Bhaduri \textit{et al.}, 
J. Phys. A: Math. Gen. 25, 6163 (1992). 

\bibitem{Lee-Fisher:89 & Moore-Read:91}
B. I. Halperin, Phys. Rev. Lett. 52, 1583 (1984);
D. Arovas, J. R. Schrieffer, F. Wilczek, \textit{ibid.} 
53, 722 (1984); 
D.-H. Lee, P. A. Fisher, 
\textit{ibid.} 
63, 903 (1989); 
G. Moore, N. Read, 
Nucl. Phys. B 360, 362 (1991); 
G. S. Jeon, K. L. Graham, J. K. Jain, Phys. Rev. Lett. 91, 036801 (2003);
E. J. Bergholtz \textit{et al.}, \textit{ibid.} 99, 256803 (2007).

\bibitem{Goldin-Sharp:96 & Goldin-Majid:04}
A. Lerda, S. Sciuto, Nucl. Phys. B 401, 613 (1993);
G. A. Goldin, D. H. Sharp, 
Phys. Rev. Lett. 76 1183, (1996); 
G. A. Goldin, S. Majid, 
J. Math. Phys. 45, 3770 (2004). 

\end{thebibliography}
\end{document}